# Enforcing Semantic Integrity on Untrusted Clients in Networked Virtual Environments[*]

Uwe Hermann, Stefan Katzenbeisser, Christian Schallhart, Helmut Veith
Technische Universität München
Institut für Informatik (I7)
Boltzmannstr. 3, D-85748 Garching
{hermanuw,katzenbe,schallha,veith}@in.tum.de

September 25, 2018

## Abstract

During the last years, large-scale simulations of realistic physical environments which support the interaction of multiple participants over the Internet have become increasingly available and economically significant, most notably in the computer gaming industry. Such systems, commonly called networked virtual environments (NVEs), are usually based on a client-server architecture where for performance reasons and bandwidth restrictions, the simulation is partially deferred to the clients. This inevitable architectural choice renders the simulation vulnerable to attacks against the semantic integrity of the simulation: malicious clients may attempt to compromise the physical and logical laws governing the simulation, or to alter the causality of events a posteriori.

In this paper, we initiate the systematic study of semantic integrity in NVEs from a security point of view. We argue that naive policies to enforce semantic integrity involve intolerable network load, and are therefore not practically feasible. We present a new semantic integrity protocol based on cryptographic primitives which enables the server system to audit the local computations of the clients on demand. Our approach facilitates low network and CPU load, incurs reasonable engineering overhead, and maximally decouples the auditing process from the soft real time constraints of the simulation.

## 1 Introduction

Interactive physically plausible simulations of artificial worlds have evolved from a topic of philosophical speculation and science fiction into an increasingly important and highly active area of computer science. With the cheap availability of extensive computing power as well as networking bandwidth, today's technology has in fact surprisingly quickly reached the point of actually implementing *networked virtual environments* [20, 17] with reasonable quality. One of the major driving forces in this development has been the computer gaming industry which has successfully developed massively multiplayer online games (MMORGs) where thousands of human participants interact over the Internet in artificial environments with almost photo-realistic simulation quality [2, 1]. Gaming applications are the ideal playground for advancing NVE technology: MMORG failure scenarios are relatively uncritical (in comparison to, say military simulations), technical features are at times more valued by the customers than total reliability, and, last not least, there are large economic interests from major industrial players including Microsoft, IBM and Sony.

The long term impact of NVE technology, however, will be a paradigm switch whose impact reaches far beyond computer gaming: it is widely expected that advanced Internet services facilitating online collaboration, e-commerce, virtual communities, entertainment, medicine, design, etc. will be based on NVE technology. Recent special issues by the IEEE Communications Magazine [5] on NVE technology as well as Communications of the ACM [3, 4] (on the transition of gaming technology into canonical computer science, and on interactive immersion in 3D graphics) reflect the rapidly increasing industrial interest in NVEs.

[*]The work described in this paper has been supported in part by the European Commission through the IST Programme under Contract IST-2002-507932 ECRYPT.

Despite the evident industrial interest in NVE technology, academic contributions to this emerging field hitherto have been mostly confined to the virtual reality community. The technical issues arising in NVEs however pose important challenges to the state of the art in many classical areas of computer science, including software engineering, distributed systems, databases, and also computer security. The goal of the current paper is to *initiate the systematic study of security issues in* NVEs *and to present security protocols which prevent malicious participants from compromising the semantic integrity of the NVE.*

In order to put our technical results in perspective, we will first review the architecture of NVEs.

**NVE Architecture.** Following [20], a virtual environment is an interactive, immersive, multi-sensory, 3 dimensional, synthetic environment. Typically, the participants of a virtual environment are represented by avatars (i.e., virtual characters) that can move through a realistic virtual world inhabited by various computer-controlled active objects. When the virtual environment is distributed over a number of hosts, we speak of a networked virtual environment (NVE). We can naturally distinguish two crucial purposes of networking in NVEs:

- **Clustering:** Since the computational load of a large-scale virtual environment is enormous, the simulation needs to be distributed over a number of computers. To this aim, a relatively small number of nodes is interconnected within an enclosed cluster. To the outside world, such a cluster ideally mimics the behavior of a single machine (which we shall call StateServer later on.) Since all resources within the cluster are solely dedicated to the NVE, we will assume that they are mutually trusted. Consequently, clustering does not directly affect the security of the NVE.

- **Remote Access:** Networking is an obvious necessity as soon as a number of geographically dispersed persons or client computers have to interact within the same virtual environment. In this case however, neither the connection to the server system *nor the client itself* are under control of the NVE system. Consequently, NVEs with remote access not only have to cope with the deficits of the underlying networking infrastructure (long transmission times and frequent packet loss, see [18] for an overview), but also with *malicious clients attacking the* NVE .

There are two principal architectures to enable remote access, namely peer-to-peer and client-server. In this paper we are *only concerned with remote access* NVEs *which are based on the more important client-server model* [12]. In a client-server NVE architecture, the authoritative and central version of the state of the NVE is maintained by the server system StateServer. The clients $\mathsf{Client}_1, \ldots, \mathsf{Client}_n$ have private and limited access to the central state of the NVE.

Each time $\mathsf{Client}_i$ wants to update the shared state of the NVE, it has to send a corresponding request to StateServer. StateServer checks whether the requested actions are compliant to the rules of the NVE. If this is the case, StateServer sends to $\mathsf{Client}_i$ an authoritative state update message that contains (as acknowledgement) the requested state update of $\mathsf{Client}_i$ as well as all changes to the central state that occurred since the last update message was sent to $\mathsf{Client}_i$. Finally, $\mathsf{Client}_i$ updates its local state according to the answer received from the server system.

By a *client cycle* we shall understand one complete turnaround of the client state as described above, i.e., the compound procedure which starts with the computation of the update request by $\mathsf{Client}_i$ and ends with the client state update according to the server response.

**Security and Semantic Integrity.** In NVEs facilitating remote access, a large number of potentially anonymous participants interact with the server using client software which can be modified by malicious participants. Consequently, any security assessment of NVEs must assume that all clients are untrusted. In Section 2, we provide a systematic taxonomy of attacks against NVEs. We argue that in addition to the classical security issues of networked systems, NVEs can be affected by attacks against their *semantic integrity*: A malicious client may attempt to compromise the physical and logical laws governing the simulation, or to alter the causality of events a posteriori. Such attacks have been repeatedly reported for MMORG game applications, yet are rarely found in the academic literature. Semantic attacks evidently represent enormous risks for the commercial NVE applications mentioned above.

Instead of implementing the centralized simulation model outlined previously, NVEs typically employ a distributed simulation model where the server only maintains a simplified ("abstract") global state, whereas the detailed simulation (e.g., graphical rendering, or extrapolation of events) is partially deferred to the clients. Such an approach reduces the processing power requirements for updating the state maintained at the server and reduces the bandwidth necessary to exchange state updates. However, the *semantic gap* between the server and the clients can be exploited by malicious clients. Such a client can successfully submit *spurious updates*, i.e., updates that are consistent with the rules of the NVE on the abstract level, but violate the NVE semantics on the concrete level.

For example, in many entertainment NVEs, the server system is only maintaining rough geometric representations of physical objects in the simulated world. Consequently, the server-side collision detection is much more efficient but is also more permissive, while the clients are responsible for implementing the collision detection at a detailed level. A malicious client can exploit this situation by moving an object along a path that is correct with respect to the rough abstract state but is incorrect with respect to the concrete state. Figure 1 illustrates this situation with a simple example where an object has to be moved through a tunnel: Since the server system maintains an abstracted geometrically simple view of the world (where the boundaries in this case are obtained from convex hulls of certain anchor points), the client is able to claim that the object can actually be moved through the tunnel. Since the inconsistent update will be accepted by the server, the malicious client has successfully broken the rules of the NVE.

Although the given geometric example may appear simplistic, every reasonably powerful abstraction will inevitably result in information loss, enabling malicious clients to submit spurious updates.

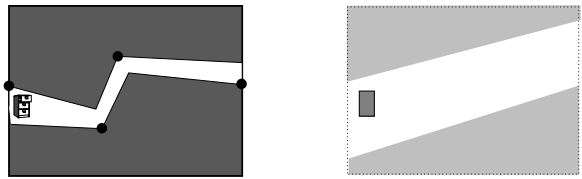

Figure 1: Example of a semantic attack.

**Engineering requirements.** Performance and scalability constraints require large-scale NVEs to employ an architecture where significant computations are performed at the client side; see [17] for a comprehensive overview. More specifically, any NVE implementation should satisfy the following two requirements:

- **Autonomous Clients:** The client of an NVE must be autonomous to provide a fluent simulation. In particular, the client has to extrapolate missing information and to interpolate between discrete client cycles. For example, the client is responsible for extrapolating the current position of an object if a position update is lost due to the deficits of the network connection.

- **Abstracted Central State:** As argued above, StateServer can only maintain the abstracted information which is necessary to coordinate the clients, while the clients utilize a more concrete version of the state. In general, the state server has a more "abstract" view of the world than the clients.

A useful *security solution* for NVEs must be designed to meet the following requirements in addition:

- **Minimal and Scalable Resource Overhead:** The overhead caused by the security solution should be as small as possible and scale well with respect to the number of participants and the level of security to be enforced. While this sounds like an obvious requirement, we want to stress that NVEs produce heavy CPU and network loads. *Therefore, any approach* *which causes significant CPU or network overhead at the client- or sever-side is impractical.*

- **Minimal Quality of Service Requirements.** A security solution should utilize unreliable network services for as many messages as possible. Only a few important messages should require timely and reliable send operations.

- **Minimal Engineering Overhead:** NVEs are complex software systems which usually consist of multiple large software packages. Thus, it is often infeasible to modify an existing NVE to meet strict security requirements. Because of the growing legacy code which is used to build new NVEs, *a security solution should be as independent as possible from the simulation and should only affect small portions of the code.*

**Technical contribution.** The main technical contribution of this paper is a set of protocols which allow to maintain the states of the clients and the server consistently and securely, even in the presence of maliciously modified clients. The approach is based on an efficient audit procedure that is performed repeatedly and randomly on the NVE clients. During the audit process, it is verified whether the *concrete* state updates performed by the client in a specific time frame are valid according to the NVE semantics. The solution meets all above stated requirements:

- The protocols enforce semantic integrity on the NVE clients, while allowing a central abstracted state and autonomous clients.

- The solution incurs very low additional network traffic, and requires the transmission of complete client states over the network *only during the audit process*. More precisely, our solution requires only a few additional bytes per client cycle which is a negligible quantity in comparison to the other messages in the client cycle. The security overhead consists solely of a Messages Authentication Code which is considered secure at a size of 16 bytes. On the other hand, a modern MMORG requires roughly one kilobyte per client and cycle. Note that with thousands of clients, bandwidth is a bottleneck mainly at the server side.

- Our solution uses reliable and time critical network transmissions only for a few small messages. All other messages, in particular the complete audit process, can be implemented solely using unreliable send operations.

- The audit process is completely independent of the (time critical) simulation, and will in general not affect the smoothness of the simulation for all but possibly the audited party.

- The protocol is designed as to be integrated into existing middleware and clients with tolerable overhead.

**Related Work on Security.** Audit trails were successfully applied in electronic commerce applications (e.g., see [13]). An audit trail enables a special party, called *auditor*, to verify the correctness of previous transactions. The audit trail can either be stored at the client or the server. In any case, the audit information must be protected from modifications. Bellare and Yee [8] identified *forward security* as the key security property for audit trails: even if an attacker completely compromises the auditing system, he should not be able to forge audit information referring to the past. Implementations of secure audit and logging facilities can be found in [16, 15, 9].

The protocols described in this paper follow the principles of audit trails, but account for the specific particularities of NVE environments. Most importantly, our solution incurs a minimal network traffic overhead, while retaining its security. In fact, a direct adoption of classical audit trails to the NVE scenario would inflict a large load on the network, as the concrete state updates of all clients must be verified. In our solution, the audit information is stored at the client side and sent to the auditor on request. The client only "commits" itself to a status update by sending a short message to the server, which cannot be altered later.

Only a few papers have dealt directly with security in online games. Baughman and Levine [7] concentrated on peer to peer multiplayer games, while we consider client server architectures of large-scale online games. Yan and Choi [21, 22] gave a taxonomy of security issues in online games and a case study on security of online bridge gaming. Davis [11] points out the importance of security in online games from a business perspective.

The rest of this paper is organized as follows: In Section 2 we provide a systematic threat analysis for NVEs, whereas Section 3 provides a description of the basic state update mechanism performed by traditional NVEs which maintain a central abstracted state. Section 4 provides a detailed technical description of our proposed security solution. Finally, a summary and outlook on future research can be found in Section 5.

## 2 Threat Analysis

In the first NVEs to be deployed, the users belonged to well-defined groups whose NVE-clients were trusted. However, as large-scale NVEs with untrusted and dispersed participants are becoming more popular, the security of NVEs becomes an eminent issue. The security threats occurring in the context of an NVE with untrusted participants can be classified as follows:

1. **System Security Attacks:** There are a number of classical security problems associated with NVEs, such as authentication, payment, or host security. These security issues have been widely studied [6, 19].

2. **Semantic Subversion:** The participants of an NVE can interact in the virtual environment according to a set of rules. The enforcement of these rules is of crucial importance for all honest participants. We call attacks targeted at circumventing or subverting these rules semantic attacks.

    (a) **Semantic Integrity Violation:** Attacks in this category attempt to violate the physical and logical laws of the NVE without detection by the server. All attacks in this class involve maliciously modified software on the client side and come in two flavors:

      i. **Rule Corruption:** The malicious client attempts to modify the simulation in a way that is illegal but plausible to the server system.
      ii. **Causality Alteration:** The malicious client attempts to withdraw previous actions to obtain unfair advantages, i.e., the client attempts to "rewrite its history".

    (b) **Client Amplification:** In this case, the client employs special software to achieve capabilities to exploit the possibilities of the NVE in an unintended manner. During such an attack, the externally observable behavior of the amplified client is not reliably distinguishable from the behavior of a honest client. Amplification attacks contain the following two main categories:

      i. **Sniffing:** The malicious client exposes information which has to be downloaded for technical reasons but is not intended to be observable immediately.
      ii. **Agents:** The malicious client enhances the natural capabilities of the human participant, e.g., by logging and evaluating previous events systematically, or by partially replacing the human participant with automated search strategies.

3. **Metastrategies:** Attacks in this category are compliant with the NVE and do not involve software modifications. They exploit principal vulnerabilities present in the NVE, e.g., collusive collaboration of human participants, or mobbing of fellow participants.

Note that system security attacks are targeted against the server systems, while all other attack groups identified in this section describe exploits which involve the client side only.

System security attacks are exploits that do not involve specific properties of NVEs and therefore they are not in the scope of this paper. On the other extreme, Metastrategies cannot be coped with by technical means. Consequently, the focus of this paper is on Semantic Subversion Attacks; these attacks are further subdivided into the categories Semantic Integrity Violation and Client Amplification. The latter cannot be handled in a rigorous way, but are amenable to statistical detection and countermeasures similar to intrusion detection systems. Thus, we have identified **Semantic Integrity Violation** as the main NVE-specific class of attacks which needs to be treated at the protocol level. The protocols presented in this paper consider both rule corruption and causality alteration attacks. To do so, the protocols enforce the following two conditions on the client behavior:

- **Rule Compliance:** Each client is only allowed to act in accordance with the rules of the NVE. This prevents rule corruption.

- **Action Commitments:** Critical actions initiated by the client must be executed in an unrevocable and undeniable manner. Consequently, clients are not allowed to choose an alternative history of actions once they obtain more information in the future. This condition prevents causality alteration.

## 3 Unsecured Client Cycle

In this section, we review the state update mechanism that is commonly implemented in NVEs that maintain a central abstract state. We write ASTATE to denote the centrally maintained and abstracted state. The portion of ASTATE which is accessible to $\text{Client}_i$ is denoted by $\text{ASTATE}[\text{Client}_i]$ (this portion depends on the spatial position of $\text{Client}_i$ in the simulated world). Clients have to map these abstracted states to concrete ones. Given an abstract state $s$, we use $\gamma(s)$ to denote the set of possible concretizations. Furthermore, if $S$ is a concrete state, then $\alpha(S)$ is the unique abstract state which corresponds to $S$. The pair $\alpha()/\gamma()$ can be naturally viewed as a Galois connection between the set of abstract and concrete states [10], i.e., $S \in \gamma(\alpha(S))$ and $s = \alpha(X)$ for any $X \in \gamma(s)$.

When connecting to the NVE, $\text{Client}_i$ receives a concrete state $S \in \gamma(\text{ASTATE}[\text{Client}_i])$ to initialize its local state $\text{STATE}[\text{Client}_i]$. From this point on, $\text{Client}_i$ maintains and updates $\text{STATE}[\text{Client}_i]$ locally.

If $\text{Client}_i$ wishes to change its state, it has to inform the StateServer in order to update the central NVE-state ASTATE. For this purpose, $\text{Client}_i$ computes a state update in the form of a compact description $\Delta$ of the difference between the current state $\text{STATE}[\text{Client}_i]_t$ and the intended next state; we call $\Delta$ a diff. Given a state $S$ and a diff $\Delta$ between $S$ and $S'$, we denote the application of $\Delta$ to $S$ by $S' = S \boxplus \Delta$. Note that $\Delta$ will typically be small compared to the state descriptions $S$ if the NVE performs a fine-grained simulation of the virtual world.

In the following, we will apply $\alpha()$ and $\gamma()$ not only to states, but also to diffs. In particular, we use $\alpha(\Delta)$ to denote the abstraction of a diff. If $S' = S \boxplus \Delta$ holds, then we require that $\alpha(S') = \alpha(S) \boxplus \alpha(\Delta)$ is also true. Furthermore, we use $\gamma(S, \delta)$ to denote the concretization of an abstract diff $\delta$ relative to a concrete state $S$. More precisely, if $S' = S \boxplus \Delta$, then $\Delta \in \gamma(S, \alpha(\Delta))$ and for all $\Delta' \in \gamma(S, \alpha(\Delta))$, we get $\alpha(S \boxplus \Delta') = \alpha(S')$.

One client cycle consists of the following steps: The client sends an abstraction of $\Delta$, denoted by $\delta = \alpha(\Delta)$, to StateServer, which evaluates the semantics of the update. Now, two cases can happen:

- If $\delta$ is allowed with respect to the semantics of the NVE, then StateServer responds with a $\delta'$ that contains all changes intended by $\text{Client}_i$ together with state updates performed by other clients present in the NVE. Upon receipt of $\delta'$, $\text{Client}_i$ computes a concretization $\Delta' \in \gamma(\text{STATE}[\text{Client}_i], \delta')$ and updates its own state by computing $\text{STATE}[\text{Client}_i]_{t+1} = \text{STATE}[\text{Client}_i]_t \boxplus \Delta'$.

- If $\delta$ is not consistent with the semantics of the NVE, the server refuses to apply the update. This can happen if $\text{Client}_i$ tries to do something impossible, such as opening a locked door. Moreover, inconsistencies can be caused by synchronization and communication errors (e.g., packet loss of the underlying network). In this case, the StateServer responds with a state update $\delta'$ which only contains the states updates of other clients.

If the clients behave according to the NVE specification, this protocol suffices to consistently maintain both the state of the clients and the server. However, if malicious clients participate in the simulation, this protocol is susceptible to *semantic integrity* violation as described in Section 1: StateServer is only able to check whether the *abstract* state updates $\delta = \alpha(\Delta)$ are consistent with its *abstract* state. A malicious Client can make an inconsistent state change $\Delta$ whose abstraction $\delta$ is consistent with the NVE rules.

In the next section, we show how to amend the basic state update protocol described above with cryptographic mechanisms in order to prevent semantic integrity violation attacks.

# 4 A Secure Semantic Integrity Protocol

In this section, we describe a protocol to enforce semantic integrity, which satisfies all requirements established in Section 1. Our approach is based on an audit procedure, which is performed by a dedicated server, namely the AuditServer. During each client cycle, the client sends a piece of evidence (containing a MAC of the applied *concrete* state update) as action commitment to AuditServer. Note that our security model assumes that AuditServer is fully trusted, which implies that a client is unable to alter past action commitments. When auditing is required, AuditServer asks a Client for a sequence of concrete state updates for a specific time frame. Based on this information, AuditServer simulates the requested segment of the Client computation and checks its compliance to the NVE rules. In addition, AuditServer checks whether the received concrete state updates match the action commitments received in the past. If both checks pass, the client is considered honest.

Audits are initiated according to a strategy determined by the server, which is unpredictable for the client. For example AuditServer might choose clients for auditing in a completely random fashion or audit "on demand" whenever statistical evidence suggests cheating. However, it is important for the security of the NVE that the clients cannot predict the time when AuditServer invokes the (next) audit protocol.

The auditing process is organized in terms of *audit cycles*, where each audit cycle consists of exactly $l$ client cycles. At each $l$th client cycle, a new audit cycle is started. At the beginning of each audit cycle, the client sends a MAC of the concrete full state as action commitment to AuditServer. As this MAC may be costly to compute because of the large state description, this message has to arrive only within the current audit cycle (i.e., within the next $l$ client cycles). In addition, as noted above, the client sends an action commitment of the applied concrete diff during each client cycle; as the diff is usually small, we require that this message arrives at AuditServer during the same client cycle.

In this paper, we assume for simplicity that $l$ is a system-wide announced and agreed on parameter. However, it is possible to customize $l$ for each client while the simulation is running. The parameter $l$ essentially determines how far back into the past auditing is possible. Consequently, the probability of successful cheating decreases with both the parameter $l$ and the frequency of the audit procedure. In particular, we shall see that there is a trade-off between the probability for successful cheating and the required additional network bandwidth.

While StateServer only keeps the current abstracted central state, the clients do not only maintain their current concrete state but also *retain a history of previous states in a local buffer,* containing up to 3 full states and $3l$ diffs. During the audit process, AuditServer requires Client to prove that its actions during the last two completely finished audit cycles and the current audit cycle are compliant to the rules of the NVE. To do so, the Client has to retain a copy of the complete state at the beginning of each new audit cycle together with diffs between the states of intermediate client cycles. All buffer content older than three audit cycles on the client side can be deleted safely.

More precisely, the buffer describes a sliding window which contains the state history of the last $2l+1$ to $3l$ client cycles, i.e., the last two full audit cycles and the current one. The sliding window which is maintained at client cycle $t_0 \geq 2l$ contains the states $S_{t_a}, S_{t_a+l}, S_{t_0}$ as well as all the intermediate diffs $\Delta'_{t_a+1}, \ldots, \Delta'_{t_0}$ where

$$t_a = \left\lfloor \frac{t_0}{l} - 2 \right\rfloor l. \quad (1)$$

Thus $t_a$ denotes the expiration time for client side audit information. In addition to the history of states, the client stores all messages received from the server within the time interval determined by the sliding window.

Figures 2a, b, and c show the gradual change of the buffer of one specific client. These figures illustrate the buffer contents at client cycles $7l+1$, $7l+2$, and $7l+3 = 8l$, respectively. The symbol ● represents a fully saved state, whereas ▲ represents a concrete diff, both saved at the client. On the other hand, ○ and △ represent action commitments of full states and diffs which are available at the AuditServer.

As seen in Figure 2, at most three fully saved states are retained at any given time; the scope of an audit process covers at most three audit cycles (see Figure 2b). Once a new audit cycle is completed, the information about an earlier audit cycle can be discarded (see Figure 2c).

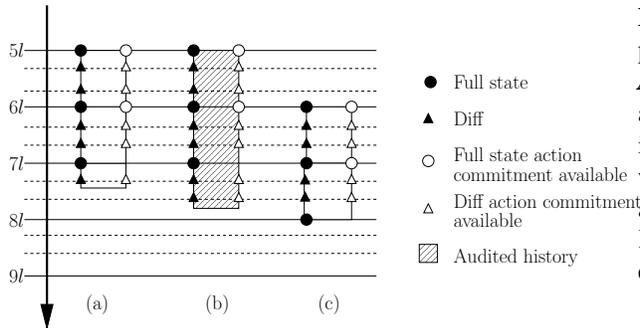

Figure 2: The "sliding window".

Crucial to the correctness of the audit process is the enforcement of the timing conditions for the action commitments. The action commitment of a diff *must* arrive within the current client cycle, whereas action commitments of full states must be available only when the current audit cycle is completed. In Figure 2 the action commitments (represented by $\triangle$) for diffs are available at the AuditServer immediately. In contrast, the action commitment $\circ$ for the full state $7l$ becomes available when the system enters state $8l$.

Note that the late availability of the full state action commitment messages requires the audit process to audit at least two full audit cycles, as otherwise the MAC (and thus also the semantic integrity) of the intermediate state $S_{t_{a+l}}$, which serves as possible future audit starting point, cannot be verified. In contrast, the protocols can easily be adapted in such a way that more audit cycles are verified during each invocation of the audit protocol. However, for the sake of brevity, we present the protocols for the simplest case of auditing at most three audit cycles.

If Client is audited, it has to send the contents of the current sliding window (i.e., the state information and corresponding state server messages) to AuditServer. First, AuditServer checks whether the received state information machetes the action commitments received previously. Second, AuditServer checks whether the client computation is compliant to the rules of the NVE by simulating the client. The audit results in a positive verdict if and only if both checks succeed.

**Protocol Description.** Secure integrity enforcement is performed by three protocols *Initialize*, *StatusUpdate* and *Audit*. The protocol *Initialize* is performed whenever a client wants to join the NVE, whereas *StatusUpdate* is executed at each client cycle (i.e., whenever a client wishes to change its state). Finally, *Audit* implements the auditing mechanism. We assume that a client leaving the NVE performs an ordinary status update, where the diff encodes the intention to leave the NVE.

For the sake of simplicity, we present the protocol for a single client Client that interacts with StateServer and AuditServer. For multiple clients, the protocol is processed asynchronously in parallel. Sending a message unreliably will be denoted by $\hookrightarrow$. Sending a message reliably that must arrive before the next $t$-th client cycle is initiated, will be denoted by $\leadsto_t$. Unreliable messages may be dropped or delivered with delay. However, we assume that no packet corruption occurs.

In the protocols we use a Message Authentication Code (MAC) as cryptographic primitive. The tagging algorithm, which takes a message $m$ and a key $k$ and produces a MAC $t$, is denoted by $t = \text{MAC}(k, m)$, whereas the verification algorithm is written as $Verify(k, m, t) = \{\text{TRUE}, \text{FALSE}\}$. We write $M = AuthMsg(m, \text{Client}, k)$ as an abbreviation for $m \parallel \text{MAC}(k, m \parallel \text{Client})$, where $\parallel$ denotes string concatenation. Furthermore, we will denote with $M^{(1)}$ and $M^{(2)}$ the two parts of the message $M$, i.e., $M^{(1)} = m$ and $M^{(2)} = \text{MAC}(k, m \parallel \text{Client})$. For the sake of simplicity we will abbreviate $\text{STATE}[\text{Client}]_t$ with $S_t$.

The protocols use two different MAC keys: $k_{\text{StateServer}}$ is mutually agreed between the state server and the audit server and is used to authenticate status updates sent from StateServer to Client. The MAC enables AuditServer to check whether a cheating Client has passed modified status update messages to the AuditServer. The key $k_{\text{Client}}$ is agreed between Client and AuditServer and is used to authenticate action commitments sent to AuditServer during each client cycle.

In the following, we describe each protocol in detail:

*Initialize*: This protocol initializes the state of a Client wishing to join the NVE (see Figure 3).

Upon opening a connection to StateServer, the client

1. Client and AuditServer exchange a MAC session key $k_{\text{Client}}$
2. Client initializes $t := 0$ and sends an initialization request to StateServer.
3. StateServer chooses $S \in \gamma(\text{ASTATE}[\text{Client}])$
4. StateServer $\hookrightarrow$ Client : $\quad M_0 := \textit{AuthMsg}(k_{\text{StateServer}}, S \parallel n_0, \text{Client})$
5. Client sets $S_0 := S$
6. Client $\rightsquigarrow_l$ AuditServer : $\quad Q_0 := \text{MAC}(k_{\text{Client}}, S_0)$

Figure 3: Protocol *Initialize*

1. Client computes a desired status change $\Delta_{t+1}$ and its abstraction $\delta_{t+1} = \alpha(\Delta_{t+1})$
2. Client $\hookrightarrow$ StateServer : $\quad \delta_{t+1}$
3. Upon receiving $\delta_{t+1}$, StateServer computes a new $\delta'_{t+1}$ and updates its central state ASTATE accordingly
4. StateServer $\hookrightarrow$ Client : $\quad M_{t+1} := \textit{AuthMsg}(k_{\text{StateServer}}, \delta'_{t+1} \parallel n_t + 1, \text{Client})$
5. Client chooses $\Delta'_{t+1} \in \gamma(S_t, \delta'_{t+1})$ and computes $S_{t+1} = S_t \boxplus \Delta'_{t+1}$
6. Client stores $\Delta'_{t+1}$
7. Client $\rightsquigarrow_1$ AuditServer : $\quad D_{t+1} := \text{MAC}(k_{\text{Client}}, \Delta'_{t+1})$
8. Client increments $t$
9. if $t \bmod l = 0$

    (a) Client deletes all $\Delta'_{t-i}$ with $2l \leq i < 3l$ as well as the full state $S_{t-3l}$ (if $t \geq 3l$).
    (b) Client stores $S_t$ and starts to compute $Q_t := \text{MAC}(k_{\text{Client}}, S_t)$.
    (c) After computation of $Q_t$, Client $\rightsquigarrow_l$ AuditServer : $\quad Q_t$.

Figure 4: Protocol *StatusUpdate*

> 1. AuditServer $\hookrightarrow$ Client :   $audit \parallel t_0$
> 2. Client computes $t_a = \lfloor \frac{t_0}{l} - 2 \rfloor l$
> 3. Client $\hookrightarrow$ AuditServer :   $S_{t_a} \parallel \Delta'_{t_a+1} \parallel \ldots \parallel \Delta'_{t_0} \parallel M_{t_a+1} \parallel \ldots \parallel M_{t_0}$
> 4. AuditServer computes $\hat{S}_{i+1} = \hat{S}_i \boxplus \Delta'_{i+1}$ for $i = t_a, \ldots, t_0 - 1$ where $\hat{S}_{t_a} = S_{t_a}$
> 5. For all $i = t_a + 1, \ldots, t_0$, AuditServer checks whether $\Delta'_i$ is chosen from $\gamma(\hat{S}_i, \delta'_i)$ compliant to the rules of the NVE, where $\delta'_i$ is taken from the message $M_i$
> 6. For all $i = t_a + 1, \ldots, t_0$, AuditServer checks whether
>    (a) *Verify*($k_{\text{StateServer}}, M_i^{(1)} \parallel \text{Client}, M_i^{(2)}$) = TRUE and
>    (b) *Verify*($k_{\text{Client}}, \Delta'_i, D_i$) = TRUE
> 7. AuditServer checks whether *Verify*($k_{\text{Client}}, S_{t_a}, Q_{t_a}$) = TRUE and *Verify*($k_{\text{Client}}, \hat{S}_{t_a+l}, Q_{t_a+l}$) = TRUE.
>    If $t_a = 0$, Client $\hookrightarrow$ AuditServer :  $M_0^{(2)}$ and AuditServer checks *Verify*($k_{\text{StateServer}}, S_0, M_0^{(2)} \parallel \text{Client}$) = TRUE.
> 8. AuditServer accepts the computations of Client if and only if all tests in steps 5 to 7 passed.

Figure 5: Protocol *Audit*

receives the relevant status information together with a randomly generated nounce $n_0$ and a MAC of the message. At this point the state server transmits a *concrete state* $S \in \gamma(\text{ASTATE}[\text{Client}])$ to the client. The client initializes its local state $S_0$ with $S$. Note that, this is the only point, besides the audit procedure, where a concrete state is transmitted. Finally, the client sends as evidence a MAC of its state $S_0$ reliably to the audit server; as the MAC of the concrete state may be costly to compute, we only require that this sending process is completed before the $l$th client cycle is initiated.

*StatusUpdate*: After initialization, the client uses this protocol to update its local state in each client cycle to reflect actions of the client itself, of other clients, and the state server. Formally, the protocol is shown in Figure 4.

Suppose the client is in state $S_t$ and wants to change its state according to the diff $\Delta_{t+1}$. To initiate the update protocol, the client sends an abstracted diff $\delta_{t+1} = \alpha(\Delta_{t+1})$ to StateServer. The server checks whether this request is valid and consistent with the current central NVE state ASTATE and computes a new update description $\delta'_{t+1}$. This update description might differ from $\delta_{t+1}$ since it has to reflect changes of other clients and the server itself; however, if $\delta_{t+1}$ is legitimate with respect to the NVE semantics, $\delta'_{t+1}$ contains the state changes of $\delta_{t+1}$. If $\delta_{t+1}$ violates the semantic integrity, $\delta'_{t+1}$ *only* contains the state updates of the other clients but *not* $\delta_{t+1}$ (or at most those actions in $\delta_{t+1}$ that are consistent). The StateServer updates its centrally managed state ASTATE according to $\delta'_{t+1}$ and sends $\delta'_{t+1}$ back to the client, together with a MAC and an incremented nounce (steps 1-4 of the protocol).

The client now computes a concrete state update $\Delta'_{t+1} \in \gamma(S_t, \delta'_{t+1})$ and applies it to $S_t$ to enter the next state $S_{t+1} = S_t \boxplus \Delta'_{t+1}$. Finally the client sends a MAC of the concrete diff $\Delta'_{t+1}$ as action commitment reliably to the AuditServer before the next client cycle is started (this message is denoted by $D_i$). Additionally, at the beginning of each audit cycle, the client sends a MAC of its full state to AuditServer; this message can be sent unreliably and must arrive within the current audit cycle (i.e., within the next $l$ client cycles); this message is denoted by $Q_i$ (steps 5-8 in the protocol).

For audit purposes, the client saves all information as evidence that is necessary for the audit server to simulate its computations. More precisely, at the beginning of each audit cycle, the client saves its full state; in intermediate client cycles, the client only retains diffs to the previous state. In addition, the client saves all messages $M_i$ received from the state server. Finally, all outdated audit information (i.e., the fully saved state, all diffs and messages belonging to the third-last audit cycle) can be removed (step 9 in the protocol).

*Audit*: During the audit protocol, AuditServer validates the computations of one Client. In particular, AuditServer checks whether the client can present concrete state updates that match the action commitments received so far and are consistent with the NVE rules; see Figure 5.

The auditing protocol is initiated by an audit message sent to the Client during client cycle $t_0$. An audit can be initiated at any client cycle $t_0 \geq 2l$. The client first computes the starting point $t_a$ of the audit according to equation (1). The client then sends the concrete state $S_{t_a}$ as well as all diffs $\Delta'_i$ and messages $M_i$ for $t_a + 1 \leq i \leq t_0$ to the AuditServer (steps 1-3 of the protocol). Finally, the audit server checks, using the action commitment messages $D_i$ and $Q_i$ submitted by the client before, whether the client adhered to the NVE semantics. In particular, the audit server checks

- whether all $\Delta'_i$ are suitable concretizations of $\delta'_i$ sent by the state server in message $M_i$ (step 5),
- whether all state server messages $M_i$ ($t_a + 1 \leq i \leq t_0$) are unmodified (step 6a) and
- whether all action commitment messages submitted by the client beforehand are valid, in particular the AuditServer checks
  - the MACs on the messages $D_i$, $t_a + 1 \leq i \leq t_0$, (step 6b) and
  - the MACs of the full states $S_{t_a}$ and $S_{t_a+l}$, contained in the messages $Q_{t_a}$ and $Q_{t_a+l}$ (step 7). Note that these messages are already available to the audit server if the timing conditions of the *StatusUpdate* protocol are enforced.

- If the first audit cycle is to be audited ($t_a = 0$), then Client is required to present $M_0^{(2)} = $ MAC($k_\text{StateServer}$, $S_0$‖Client) to AuditServer additionally to prove that the initial state $S_0$ has been authorized by the StateServer (step 7).

If all checks pass, the client is considered honest (step 8).

**Security.** Our protocols enforce rule compliance because in each client cycle the client is committing to its concrete diff $\Delta'$. In addition, the client "commits" to its concrete state $S_t$ in each $l$-th client cycle. After this information arrived at AuditServer, the client cannot cheat about its past states (in theory it is possible for a malicious client to send a MAC of an incorrect state in message $Q_t$, but this would be noticed during the audit process). The audit server is thus able to fully simulate and validate the computations of the client. The unforgeability of the MAC implies that the client cannot alter its chosen state transitions a posteriori. Furthermore, if the timing conditions are enforced (i.e., the network delivers all ⇝-messages reliably in time) the protocols do not allow causality alteration, as the MAC must arrive at the audit server within the current client cycle.

**Computational Overhead.** The protocol can be implemented in a very resource efficient manner: The *StatusUpdate* protocol requires only a few MAC computations over relatively small amounts of data. The MAC computation over the complete state of a client can be processed in background during the $l$ client cycles of an audit cycle. Moreover, only the MACs are additionally transmitted over the network.

In contrast, the *Audit* protocol is much more data intensive and involves a complete re-simulation of the client computations. However, the execution of the *Audit* protocol is not time critical and can be delegated to a specific server, namely the AuditServer. Therefore, it does not cause resource overhead at the StateServer. The degree of confidence in the computations of the clients can be enhanced by increasing the number of audit servers and thereby increasing the number of *Audit* protocol executions.

In most application scenarios, the limiting resource for increasing the probability that a client cycle is audited is the bandwidth $B$ which is available for auditing. A closer analysis shows that this probability is proportional to $\frac{B}{|\Delta|+|M|}$, where $|\Delta|$ is the maximum size of concrete diffs and $|M|$ is the maximum size of authorized state server messages. In general, the memory requirements of the client are not a limitation since the client can buffer the history on a hard disk.

To integrate our protocol into an NVE system, one has to implement the protocol logic, add the computation of the MACs at the state server and the client, and implement the audit server. It should be possible to implement the audit server by mainly reusing client code since the audit server is simulating the client computations. The biggest problem in implementing the protocol will likely be the creation of copies of the complete client state in a timely manner, as required at the beginning of each audit cycle. However, all remaining parts of the protocol can be implemented in a straight forward manner.

## 5 Conclusion and Future Work

In this paper, we have argued that networked virtual environments are an emerging network technology which has not been subject to rigorous security investigations. We have identified *semantic integrity* as a core security problem in NVEs. Untrusted and malicious clients may utilize the fact that the central NVE server can—due to the limited computing power and the deficits of the network connection—only maintain an abstracted version of the NVE state. To overcome this problem, we have introduced a new audit trail mechanism which is able to verify the compliance of the client computation. Although we allow autonomous clients, our protocols assure that regularly cheating clients will be identified with a high probability. The audit mechanism proposed in this paper can be integrated seamlessly into current NVE architectures and inflicts little engineering and resource overhead. We are currently working on formal security proofs for our protocols. The protocols presented here will be integrated into SYNEIGHT [14], a new middleware for NVEs which is developed by our group. A detailed description of the SYNEIGHT middleware will be given in future work.